\documentclass[a4,12pt]{article}

\usepackage{graphics,epsfig,subfigure}
\usepackage{graphicx}
\usepackage{color}

\newcommand{\bea}{\begin{eqnarray}}
\newcommand{\eea}{\end{eqnarray}}
\newcommand{\beq}{\begin{equation}}
\newcommand{\eeq}{\end{equation}}

\newcommand{\nn}{\nonumber}

\newcommand{\msb}{\overline{\rm{MS}}}

\def\simle{\mathrel{\rlap{\raise 0.511ex \hbox{$<$}}{\lower 0.511ex \hbox{$\sim$}}}}

\begin{document}

\begin{titlepage}

\begin{flushright}
RM3-TH/13-12
\end{flushright}
\vspace{0.5cm}

\begin{center}
  \begin{large}
    \textbf{Vacuum Insertion Approximation and the $\Delta I=1/2$ rule: \\ \vspace{0.1cm}
    a lattice QCD test of the na\"ive factorization hypothesis  \\ \vspace{0.1cm}  for $K$, $D$, $B$ and
    static mesons\unboldmath} \\
  \end{large}
\end{center}
\vspace{0.8cm}

\begin{figure}[h] \begin{center}
 \includegraphics[scale=0.30]{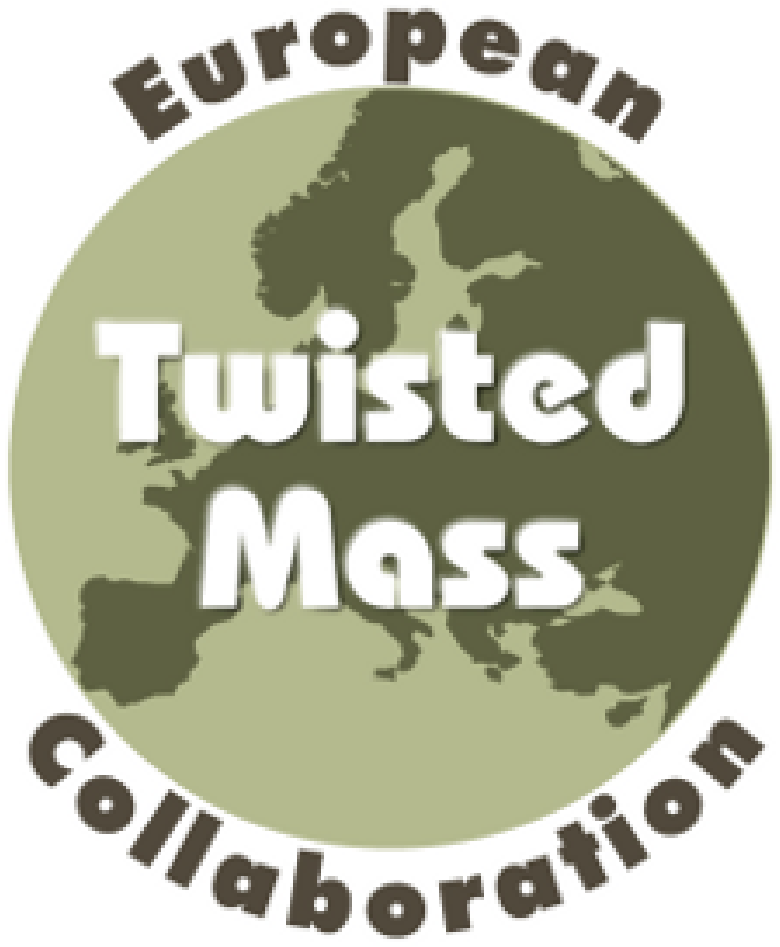}
\end{center} \end{figure}

\vspace{-0.8cm}
\begin{center}  \begin{large}
  \textbf{ N.~Carrasco$^{(a)}$, V.~Lubicz$^{(b,a)}$, L.~Silvestrini$^{(c)}$
}
 \end{large} \end{center}

\begin{center}
{\it 
$^{(a)}$ INFN, Sezione di Roma Tre \\
c/o Dipartimento di Matematica e Fisica, Universit\`a  Roma Tre \\
Via della Vasca Navale 84, I-00146 Rome, Italy \\

\vspace{0.3cm}
$^{(b)}$ Dipartimento di Matematica e Fisica, Universit\`a  Roma Tre \\
Via della Vasca Navale 84, I-00146 Rome, Italy \\

\vspace{0.3cm}
$^{(c)}$ INFN, Sezione di Roma, Piazzale A. Moro, I-00185 Rome, Italy
}
\end{center}

\vspace{0.5cm}
\begin{abstract}
  Motivated by a recent paper by the RBC-UKQCD Collaboration, which
  observes large violations of the na\"ive factorization hypothesis in
  $K \to \pi \pi$ decays, we study in this paper the accuracy of the
  Vacuum Insertion Approximation (VIA) for the matrix elements of the
  complete basis of four fermion $\Delta F=2$ operators. We perform a
  comparison between the matrix elements in QCD, evaluated on the
  lattice, and the VIA predictions. We also investigate the dependence
  on the external meson masses by computing matrix elements for $K$,
  $D_s$, $B_s$ and static mesons. In commonly used renormalization schemes, 
  we find large violations of the VIA
  in particular for one of the two relevant Wick contractions in the
  kaon sector. These deviations, however, decrease significantly as
  the meson mass increases and the VIA predictions turn out to be
  rather well verified for B-meson matrix elements and, even better,
  in the infinite mass limit.
\end{abstract}

\end{titlepage}

\section{Introduction}
A recent paper~\cite{Boyle:2012ys} by the RBC-UKQCD Collaboration
provides an emerging explanation for the ``$\Delta I =1/2$ rule" in $K
\to \pi \pi$ decays. This rule refers to the empirical observation
that the real part ${\rm Re}(A_0)$ of the amplitude describing the
kaon decay in two pions with total isospin $I=0$ is larger by
approximately a factor 22.5 than the corresponding amplitude ${\rm
  Re}(A_2)$ of the $I=2$ channel. Perturbative QCD evolution of
current-current operators from the electroweak scale down to about
$1.5-2$ GeV contributes a factor of approximately 2 to the ratio ${\rm
  Re}(A_0)/{\rm
  Re}(A_2)$~\cite{Gaillard:1974nj,Altarelli:1974exa}. Therefore,
barring significant new physics contributions to the decay amplitudes,
the remaining factor of about 10 should come from non-perturbative
QCD.

\begin{figure}[!h]	
\begin{center}
\subfigure[\it ``Connected"]{\includegraphics[width=0.33\textwidth]{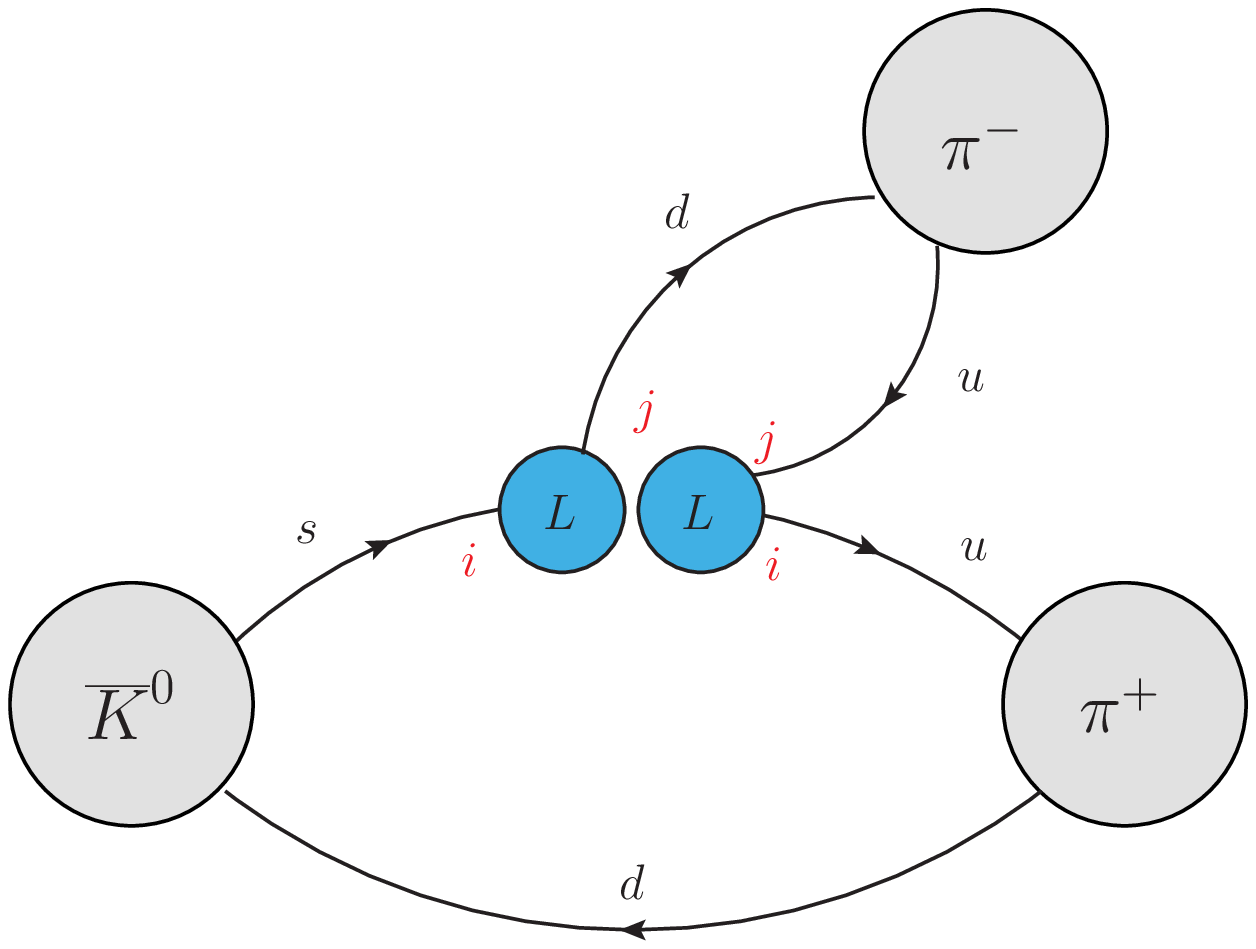}}
\subfigure[\it ``Disconnected"]{\includegraphics[width=0.30\textwidth]{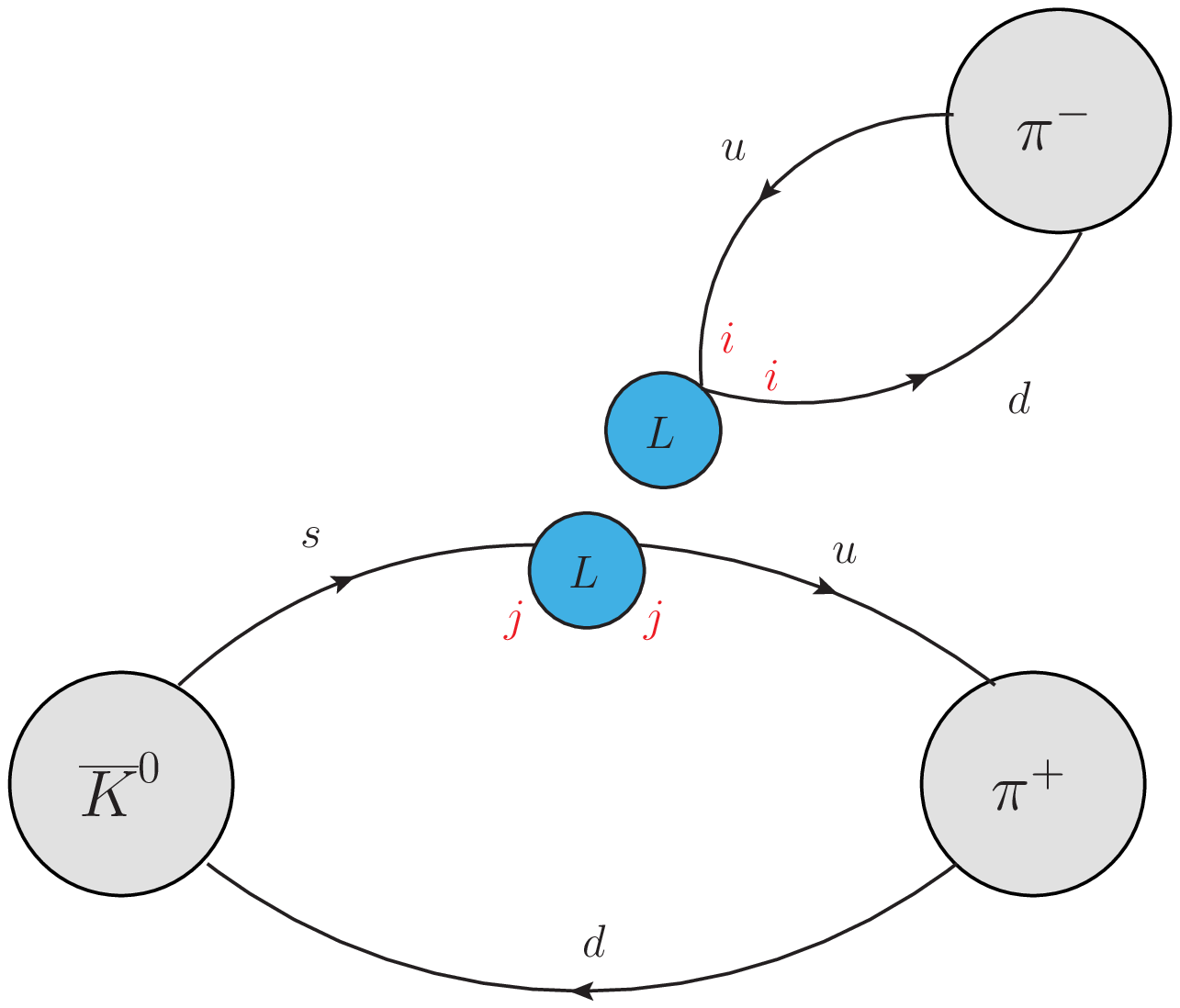}}
\end{center}
\caption{\sl ``Connected" and ``disconnected" contributions to $K \to \pi \pi$ decays in the local $V-A$ theory. The blue circles indicate the insertion of the current-current operator with left chirality. The two diagrams are distinguished by the summation of the spin (single and double trace) and color ($i$, $j$) indices.}
\label{fig:Kpipi-diagrams}
\end{figure}

The explanation of the $\Delta I =1/2$ rule which is emerging from the
lattice QCD studies~\cite{Boyle:2012ys, Lellouch:2011qw, Blum:2011pu,Blum:2011ng,Blum:2012uk} is that
the two dominant contributions to the $\Delta I =3/2$, $K \to \pi \pi$
correlation functions, which are shown diagrammatically in
Fig.~\ref{fig:Kpipi-diagrams}, have opposite signs leading to a
significant cancellation.
The same contributions are also the largest ones in ${\rm Re}(A_0)$,
but now they have the same sign and so enhance this amplitude. QCD and
electroweak penguins operators, which only enter the $\Delta I=1/2$
transition, make only very small contributions.

While the calculation of ${\rm Re}(A_2)$ by the RBC-UKQCD
Collaboration is completed, the calculation of ${\rm Re}(A_0)$ has
been only performed at unphysical kinematics, with pion masses of
about 330 MeV and 420 MeV. Therefore the results are not conclusive
yet, and the enhancement factor of 22.5 has still to be quantitatively
reproduced. Nevertheless, the emerging explanation of the $\Delta I
=1/2$ rule discussed above is rather convincing and deserves to be
further investigated.

A striking feature of the lattice results for the current-current $K
\to \pi \pi$ correlators is that they almost maximally contradict the
expectations of the na\"ive factorization hypothesis, i.e. the
predictions of the vacuum insertion approximation (VIA). Color
counting and the VIA suggest that the connected contribution of
Fig.~\ref{fig:Kpipi-diagrams} should be approximately $1/3$ of the
disconnected one, whereas it is found that in QCD they have opposite
signs. As recently stressed in ref.~\cite{Buras:2014maa}, this result
was already obtained almost thirty years ago by Bardeen, Buras and
G{\'e}rard using a model based on the dual representation of QCD as a
theory of weakly interacting mesons for large $N$ \cite{largeN}.

It is tempting to establish a connection between the validity of
na\"ive factorization for emission topologies and the $\Delta I=1/2$
rule. In the $K$ system, one observes a maximal deviation from the VIA
and a large suppression of the $\Delta I=3/2$ amplitude. In
nonleptonic charm decays, early analyses found moderate violations of
naive factorization which could be described by setting to zero the
$1/N_c$-suppressed terms in the factorized matrix elements
\cite{dnlepold}. Correspondingly, one observes comparable $\Delta
I=1/2$ and $\Delta I=3/2$ amplitudes, with a large relative phase
\cite{dmesons}. Finally, in the $B$ system factorization for emission
topologies has been demonstrated in the infinite mass limit
\cite{BBNS}, and $B \to \pi \pi$ decays can be theoretically well
described by factorization once the dominant subleading corrections
are taken into account \cite{charming}.

As already noted in Ref.~\cite{Boyle:2012ys}, a violation of similar
extent of the VIA is also exhibited by the connected and disconnected
contributions to the matrix element $\langle \overline{K}^0 \vert
(\bar s \gamma^\mu_L d)(\bar s \gamma^\mu_L d) \vert K^0 \rangle$
which contains the non-perturbative QCD effects in neutral kaon mixing
(see refs.~\cite{Buras:2014maa,Bardeen:1987vg} for a detailed
discussion of this matrix element in the context of large $N$). By
using $SU(3)$ flavour symmetry it can be shown, in fact, that the
matrix elements of the $\Delta S=2$ operator for $\overline{K}^0 -
K^0$ mixing and of the $\Delta I=3/2$ operator for $K \to \pi \pi$
decays are proportional in the soft pion limit~\cite{Donoghue}. For
this reason, earlier attempts to study $K \to \pi \pi$ decays on the
lattice were based on the evaluation of the matrix element of the
$\Delta I=3/2$ operator between a kaon and a single pion state.

\begin{figure}[!h]	
\begin{center}
\subfigure[\it ``Connected"]{\includegraphics[width=0.40\textwidth]{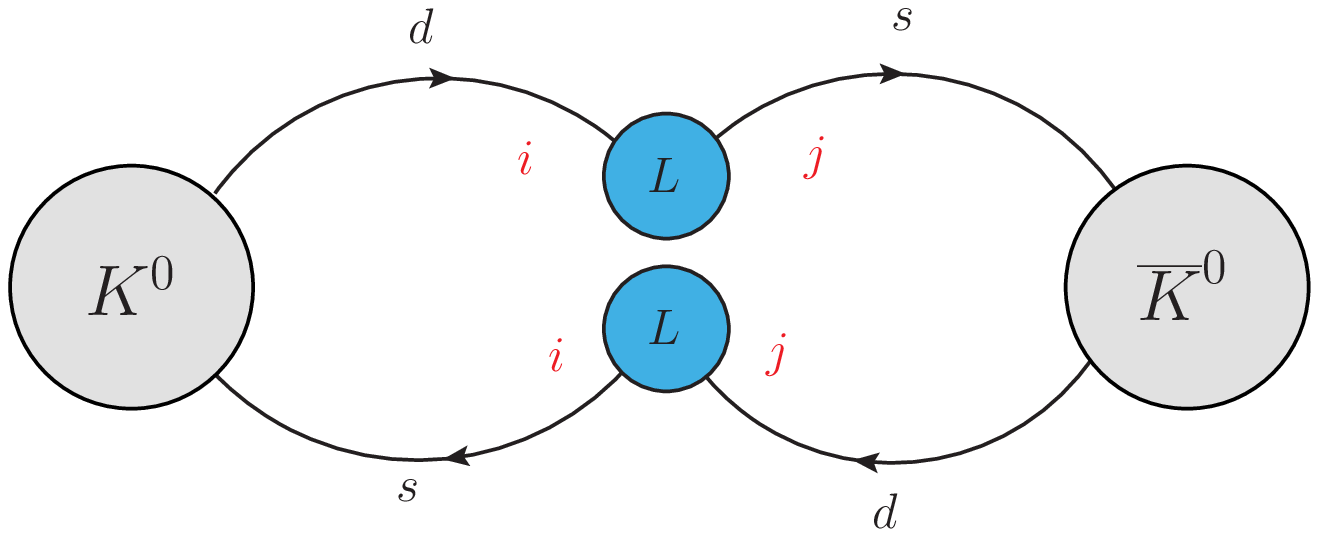}} \hspace{1cm}
\subfigure[\it ``Disconnected"]{\includegraphics[width=0.40\textwidth]{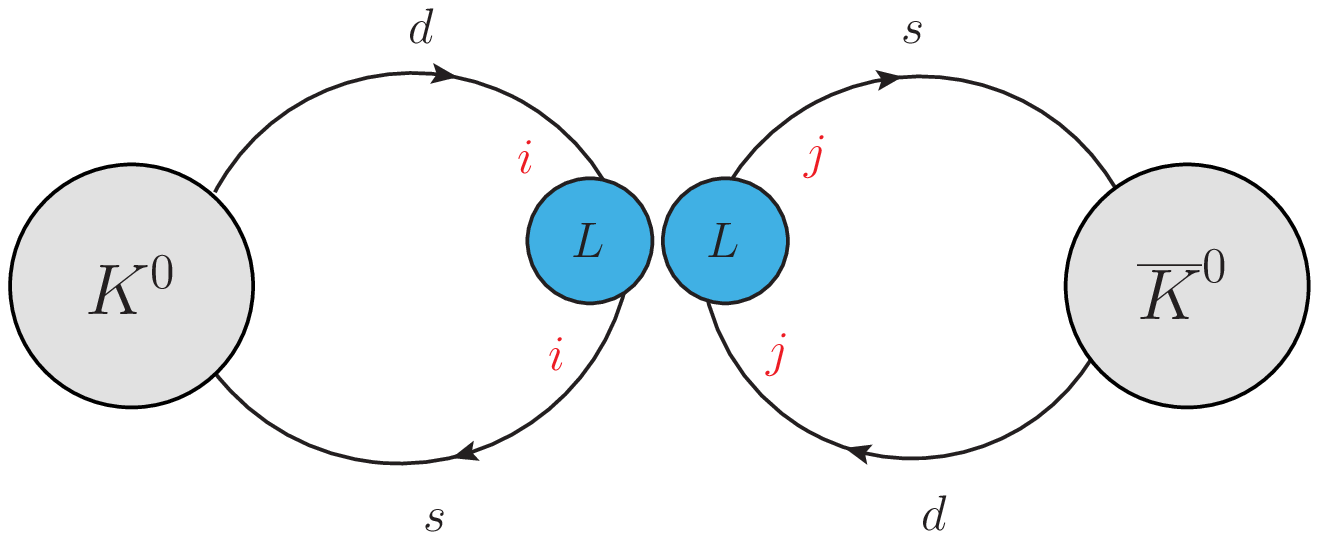}}
\end{center}
\caption{\sl ``Connected" and ``disconnected" contributions to $\overline{K}^0 - K^0$ mixing. The notation is the same as in Figure~\ref{fig:Kpipi-diagrams}.}
\label{fig:KK-diagrams}
\end{figure}

The connected and disconnected diagrams contributing to the $\langle
\overline{K}^0 \vert (\bar s \gamma^\mu_L d)(\bar s \gamma^\mu_L d)
\vert K^0 \rangle$ matrix element are shown in
Fig.~\ref{fig:KK-diagrams}.
They originate from the same Wick contractions as the analogous
diagrams for the $K \to \pi\pi$ matrix element presented in
Fig.~\ref{fig:Kpipi-diagrams}. As for the latter, the VIA predicts
that also in the $\overline{K}^0 - K^0$ case the two contributions
come in the ratio of 1/3:1, whereas the lattice calculations show that
in QCD they have opposite signs.

In this letter we further extend the comparison between QCD and VIA
predictions for the four-fermion operator matrix elements in several
respects. In particular: i) we confirm the findings of
Ref.~\cite{Boyle:2012ys} for the connected and disconnected
contributions to the matrix element of the left-left current operator
between external kaon states; ii) we extend the comparison between QCD
and VIA predictions to the whole 10-dimensional basis of four-fermion
operators, characterized by different spin and color structures; iii)
we extend the comparison to the matrix elements of heavier mesons than
the kaons, namely $D$ and $B$ mesons as well as {\it static} mesons,
i.e. mesons constituted by an heavy quark of infinite mass. Our main
results show that the VIA predictions are largely violated in QCD also
for other operators besides the left-left current operators,
particularly for the connected contributions. The discrepancies,
however, decrease significantly as the meson mass increases, and the
VIA predictions turn out to be rather well verified for $B$-meson
matrix elements and, even better, in the infinite mass limit. 
Although numerical results are presented in $\overline{MS}$ scheme
at 3 GeV, the above qualitative conclusions do not depend, to a large extent, on the 
renormalization scheme and scale. 

Our numerical results have been obtained by using the gauge
configurations generated by European Twisted Mass Collaboration (ETMC)
with $N_f = 2$ dynamical quarks at four values of the lattice
spacing~\cite{Baron:2009wt}. The 2- and 3-point correlation functions
analyzed for the present study have been computed to evaluate the
matrix elements of the four-fermion operators relevant for
$\overline{K}^0 - K^0$, $\overline{D}^0 - D^0$ and $\overline{B}^0 -
B^0$ mixing, within and beyond the Standard Model, in
Refs.~\cite{Constantinou:2010qv,Bertone:2012cu}, \cite{Carrasco:2014uya} and
\cite{Carrasco:2013zta} respectively. The results for the matrix
elements between external $B$ and static meson states have been
obtained by implementing the so called ratio method for heavy quarks
developed in Refs.~\cite{Blossier:2009hg,Dimopoulos:2011gx} and
optimized smearing techniques~\cite{Carrasco:2013zta}.

\section{Matrix elements in the VIA}
In order to study separately the connected and disconnected
contributions to the matrix element of the $\Delta F=2$ four-fermion
operator, shown in Fig.~\ref{fig:KK-diagrams} for the kaon case, we
find convenient to consider the following $\Delta F=1$ operators \bea
\label{eq:OOF}
&& O_{\Gamma\Gamma}= \left(\overline{h}\Gamma\ell\right)\left(\bar{h}'\Gamma\ell'\right) \nn \\
&& O_{\Gamma\Gamma}^{F}=
\left(\overline{h}\Gamma\ell'\right)\left(\bar{h}'\Gamma\ell\right) \
, \eea where $h,h',\ell,\ell'$ are different quark fields and $\Gamma$
is a generic Dirac matrix. It is easy to realize that the matrix
elements of the operators $O_{\Gamma\Gamma}$ and $O_{\Gamma\Gamma}^F$
between external mesons of flavour content $(\bar h \ell)$ and $(\bar
h' \ell')$ receive contribution only from the disconnected and
connected contraction respectively. In this paper, the flavours $h$
and $h'$ are always taken to be degenerate in mass, and similarly for
$\ell$ and $\ell'$. We will then present results for $(h,\ell) =
(s,d)$, $(c,s)$, and $(b,s)$ for kaon, $D_s$ and $B_s$ mesons
respectively. Moreover, we will also extrapolate the heavy quark mass
to the infinite quark mass in order to investigate the accuracy of the
VIA approximation in the static limit.

For the Dirac structure of the four-fermion operators we consider the
following (complete) basis of operators: \beq
\label{eq:basis}
O^{(F)}_{X} = \{ 
O^{(F)}_{VV+AA} \, , \
O^{(F)}_{VV-AA} \, , \
O^{(F)}_{SS-PP} \, , \
O^{(F)}_{SS+PP} \, , \
O^{(F)}_{SS+PP-TT/2} \}
\eeq
where $O^{(F)}_{VV\pm AA} = O^{(F)}_{VV} \pm O^{(F)}_{AA}, \ldots$ and
$V,\,A,\,S,\,P,\,T$ stand for $\gamma^\mu,\, \gamma^\mu\gamma^5 ,\,
1,\, \gamma^5,\, \sigma^{\mu\nu}$. With this choice, all operators
have non vanishing matrix elements in the VIA, which 
read:
\bea
&& \hspace {-0.8cm} 
\langle P^0 | O_{VV+AA} | \overline{P}'^0\rangle_{\rm VIA} =  - \langle  P^0 | O_{VV-AA} | \overline{P}'^0 \rangle_{\rm VIA} =  F^2 M^2 \\
&&  \hspace {-0.8cm} 
\langle  P^0 | O_{SS-PP} | \overline{P}'^0 \rangle_{\rm VIA} =  - \langle  P^0 | O_{SS+PP} | \overline{P}'^0 \rangle_{\rm VIA} = - \langle  P^0 | O_{SS+PP-TT/2} | \overline{P}'^0 \rangle_{\rm VIA} = \frac{F^2 M^4}{(m_h+m_\ell)^2} \nn \ ,
\eea
where $M$ and $F$ are the mass and decay constant of the pseudoscalar
$(\bar h \ell)$-meson $P^0$ and $m_{h(\ell)}$ the corresponding quark
masses. As well known, both the scheme and scale dependence of
operators and quark masses in the VIA is neglected. 

The matrix elements of the $O^{F}$ operators in the VIA are obtained
after a Fierz transformation of both spin and color indices: 
\beq
 \langle P^0 | O_{X}^F | \overline{P}'^0\rangle = \frac{1}{3} F_{XY} \langle P^0 | O_{Y} | \overline{P}'^0\rangle + \frac{1}{2} F_{XY} \langle P^0 | O^\lambda_Y | \overline{P}'^0\rangle \ ,
 \label{eq:OlambdaX}
\eeq
where  $F_{ij}$ is the Dirac Fierz matrix 
\beq
F=\left(\begin{array}{ccccc}
1 & 0 & 0 & 0 & 0\\
0 & 0 & -2 & 0 & 0\\
0 & -1/2 & 0 & 0 & 0\\
0 & 0 & 0 & 0 & -1/2\\
0 & 0 & 0 & -2 & 0
\end{array}\right)
\eeq
and the operators $O^\lambda_X$ are defined as
\beq
O^\lambda_{\Gamma \Gamma}=\left(\overline{h}\lambda^a \Gamma\ell\right)\left(\bar{h}'\lambda^a \Gamma\ell'\right)  
\eeq
with $\lambda^a$ the color group generators (Gell-Mann matrices). In the VIA one simply has:
\beq
 \langle P^0 | O^\lambda_X| \overline{P}'^0\rangle_{\rm VIA}=0 ,
\eeq
and thus, from Eq.~(\ref{eq:OlambdaX})
\bea
&& \langle P^0 | O^F_{VV+AA} | \overline{P}'^0\rangle_{\rm VIA} =   \frac{1}{3} \, F^2 M^2 \nn \\
&& \langle P^0 | O^F_{VV-AA} | \overline{P}'^0 \rangle_{\rm VIA}  = - \frac{2}{3} \,  \frac{F^2 M^4}{(m_h+m_\ell)^2} \\
&& \langle P^0 | O^F_{SS-PP} | \overline{P}'^0 \rangle_{\rm VIA}  =  \frac{1}{6} \, F^2 M^2 \nn \\
&& \langle P^0 | O^F_{SS+PP}| \overline{P}'^0 \rangle_{\rm VIA} =   \frac{1}{6} \,   \frac{F^2 M^4}{(m_h+m_\ell)^2} \nn \\
&& \langle P^0 | O^F_{SS+PP-TT/2} | \overline{P}'^0 \rangle_{\rm VIA} = \frac{2}{3} \,  \frac{F^2 M^4}{(m_h+m_\ell)^2} \nn \ .
\eea

In order to investigate the accuracy of the VIA, in the following we
will present the results in terms of ratios between the matrix
elements in QCD and their expression in the VIA, 
\beq
\label{eq:RX}
R^{(F)}_X = \frac{\langle P^0 | O^{(F)}_{X} | \overline{P}'^0\rangle}{\quad \langle P^0 | O^{(F)}_{X} | \overline{P}'^0\rangle_{\rm VIA}} \ .
\eeq
For the matrix elements of the operators $O^\lambda_X$, which vanish
in the VIA, we adopt the following normalization: 
\beq
\label{eq:Rlambda}
R^\lambda_X = \frac{\langle P^0 | O^\lambda_X | \overline{P}'^0\rangle} {\quad \langle P^0 | O_{X} | \overline{P}'^0\rangle_{\rm VIA}}\ ,\eeq
and compute the matrix elements of $O^\lambda_X$ using Eq.~(\ref{eq:OlambdaX}).
\section{Results}
The lattice calculation of the matrix elements in QCD has been
performed at four values of the lattice spacing, using the $N_f=2$
dynamical quark configurations produced by the ETM collaboration
\cite{Baron:2009wt,Boucaud:2007uk}. Quark fields are regularized by
employing the twisted mass/Oster\-wal\-der-Seiler formalism at maximal
twist, which guarantees automatic ${\cal O}(a)$-improvement and
continuum-like renormalization pattern for the four-fermion
operators~\cite{Frezzotti:2004wz}. In Table \ref{tab:runs} we provide
the main simulation details, including the values of (bare) quark
masses for each lattice spacing. The values of the light (up and down)
quark masses are equal for sea and valence quarks. We then simulate
three valence quark masses around the physical strange mass and a set
of heavy valence quark masses in the range between $m_c$ and $2.5\,
m_c$, where $m_c$ is the physical charm mass.
\begin{table}[t]
\begin{center}
\begin{tabular}{cccllc}
\hline 
{\small $a$ (fm)} & {\small $L^{3}\times T$} & {\small $N_{stat}$} & {\small $a\mu_{\ell}=a\mu_{\textrm{sea}}$} & {\small $a\mu_s$ } & {\small $a\mu_{h}$}\tabularnewline
\hline 
& {\small $24^{3}\times48$} & {\small 128} & {\small 0.0080 0.0110} & {\small 0.0175 } & {\small 0.1982 0.2331 0.2742 }\tabularnewline
{\small 0.098} &  &  &  & {\small 0.0194} & {\small 0.3225 0.3793 0.4461  }\tabularnewline
&  &  &  & {\small 0.0213} & {\small 0.5246 0.6170 }\tabularnewline
\hline 
& {\small $24^{3}\times48$} & {\small 240} & {\small 0.0040 0.0064} & {\small 0.0159} & {\small 0.1828 0.2150 0.2529 }\tabularnewline
{\small 0.085} &  &  & {\small 0.0085 0.0100} & {\small 0.0177 } & {\small 0.2974 0.3498 0.4114 }\tabularnewline
\cline{2-4} 
& {\small $32^{3}\times64$} & {\small 144} & {\small 0.0030 0.0040} & {\small 0.0195} & {\small 0.4839 0.5691 }\tabularnewline
\hline 
& {\small $32^{3}\times64$} & {\small 144} & {\small 0.0030 0.0060 } & {\small 0.0139 } & {\small 0.1572 0.1849 0.2175 }\tabularnewline
{\small 0.067} &  &  & {\small 0.0080} & {\small 0.0154} & {\small 0.2558 0.3008 0.3538 }\tabularnewline
&  &  &  & {\small 0.0169} & {\small 0.4162 0.4895 }\tabularnewline
\hline 
& {\small $32^{3}\times64$} & {\small 144} & {\small 0.0065} & {\small 0.0116 } & {\small 0.13315 0.1566 0.1842 }\tabularnewline
\cline{2-4} 
{\small 0.054} &  &  &  & {\small 0.0129 } & {\small 0.2166 0.2548 0.2997 }\tabularnewline
& {\small $48^{3}\times96$} & {\small 80} & {\small 0.0020} & {\small 0.0142} & {\small 0.3525 0.4145 }\tabularnewline
\hline 
\end{tabular}
\end{center}
\caption{\sl Details of the lattice simulations used for the present
  study. We provide the approximate value of the lattice spacing ($a$)
  \cite{Blossier:2010cr}, the number of lattice sites in the spatial
  ($L$) and temporal ($T$) directions, the number ($N_{stat}$) of
  independent gauge configurations for each ensemble, the values of
  the bare quark masses in lattice units in the light ($\mu_\ell$),
  strange ($\mu_s$) and heavy  ($\mu_h$) quark mass regions.}
\label{tab:runs}
\end{table}

\begin{figure}[!t]
\begin{center}
\includegraphics[width=0.44\textwidth]{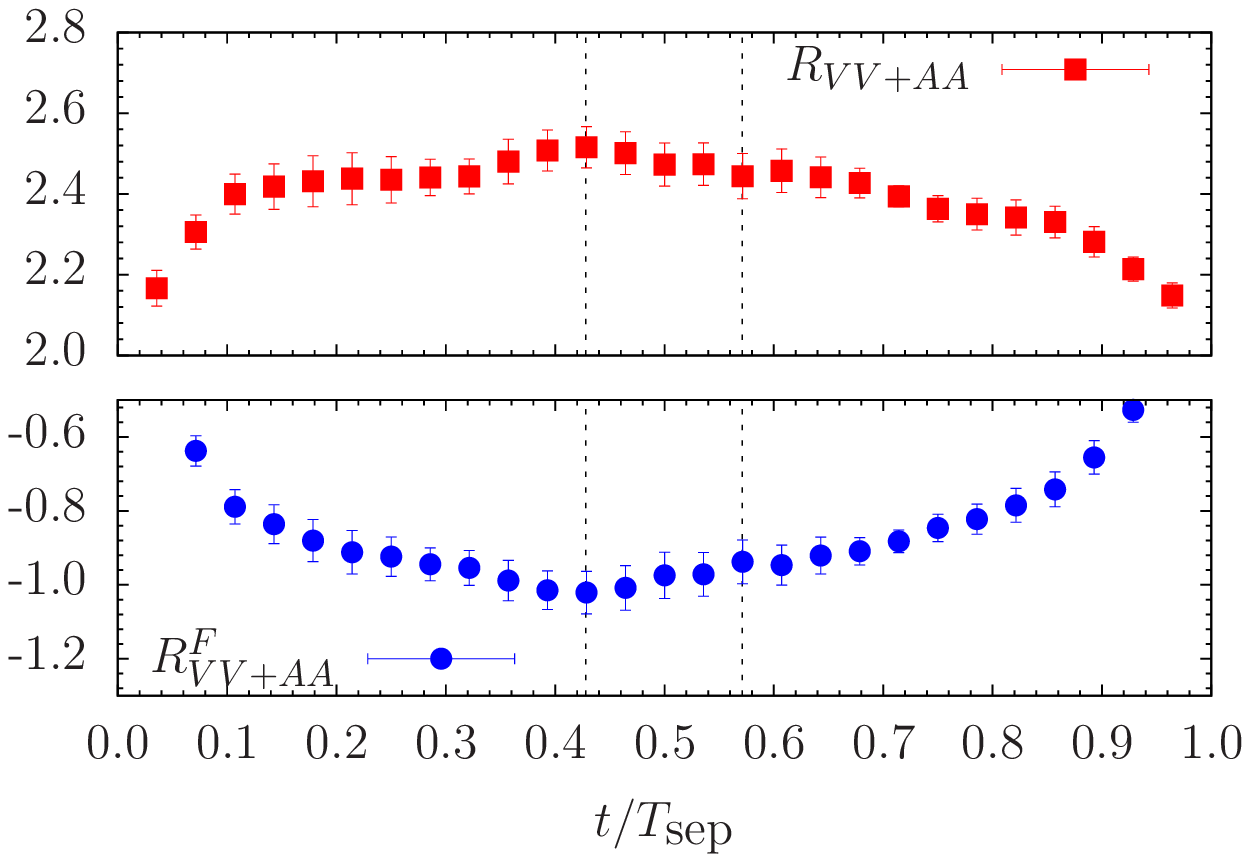} \hspace{0.5cm}
\includegraphics[width=0.40\textwidth]{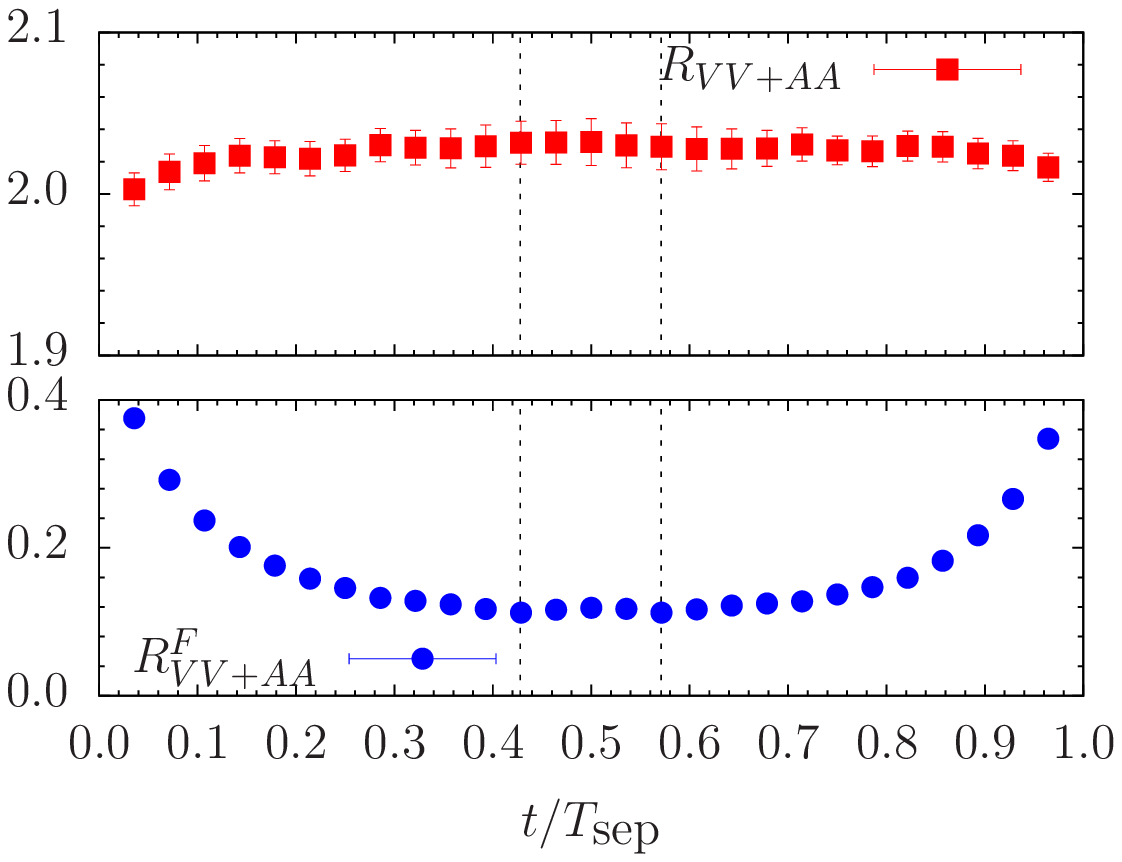}
\end{center}
\vspace{-0.5cm}
\caption{\sl Lattice data and time plateau for the estimators of
  $R_{VV+AA}^{(F)}$ as a function of $t/T_{\textrm{sep}}$, where $t$
  is the Euclidean time and $T_{\textrm{sep}}$ is the separation
  between the two external pseudoscalar meson sources. We show results
  for the finest lattice spacing ($a \simeq 0.054$ fm) and the
  lightest quark mass ($a \mu_\ell = 0.0020$). Left and right panels
  correspond to the light-strange ($K$) and strange-charm ($D_s$)
  pseudoscalar mesons respectively. The dotted lines delimit the
  plateau region from which the results for the ratios
  $R_{VV+AA}^{(F)}$ are extracted.}
\label{fig:plateau}
\end{figure}

The lattice computation of the relevant four-fermion matrix elements
proceeds as discussed in
Refs.~\cite{Constantinou:2010qv,Bertone:2012cu}, \cite{Carrasco:2014uya} and
\cite{Carrasco:2013zta} for $\overline{K}^0 - K^0$, $\overline{D}^0 -
D^0$ and $\overline{B}^0 - B^0$ mixing respectively. In the present
study, however, we evaluate separately the contributions of the
connected and disconnected diagrams and compare them with the VIA
predictions. In Fig.~\ref{fig:plateau} we show, as an example, the
lattice estimators for the ratios $R_{VV+AA} $ and $R^{(F)}_{VV+AA}$
as a function of the Euclidean time in the kaon and D-meson case. The
values of the ratios are extracted from the central region in which
the time dependent correlators exhibit a plateau.

The renormalization constants of the two- and four-fermion operators
have been computed non perturbatively in the RI-MOM scheme in
\cite{Bertone:2012cu,Constantinou:2010gr} and converted to $\msb$
using continuum perturbation theory. For renormalizing the operators
in the basis of Eq.~(\ref{eq:basis}), we simply performed a change of
basis from the four-fermion renormalization matrix reported in
\cite{Bertone:2012cu}.

\begin{figure}[!t]
\begin{center}
\includegraphics[width=0.42\textwidth]{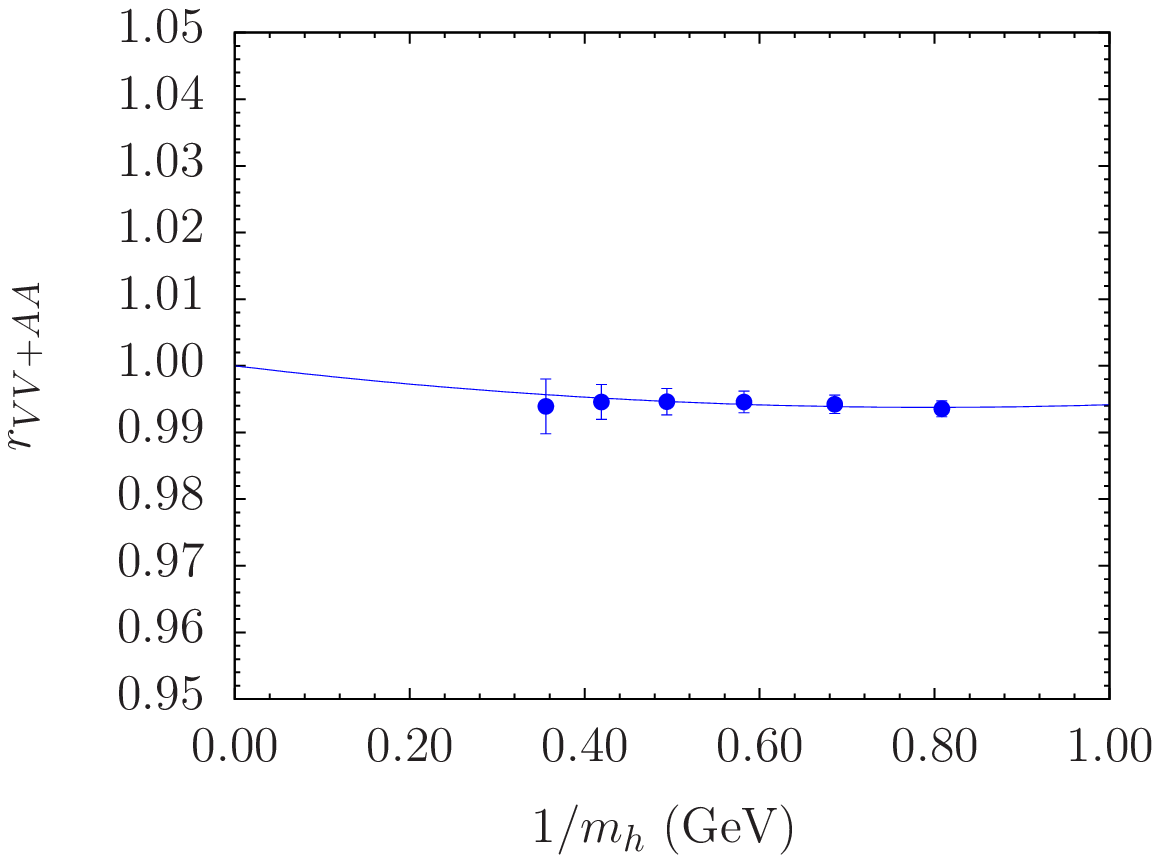} \hspace{0.5cm}
\includegraphics[width=0.42\textwidth]{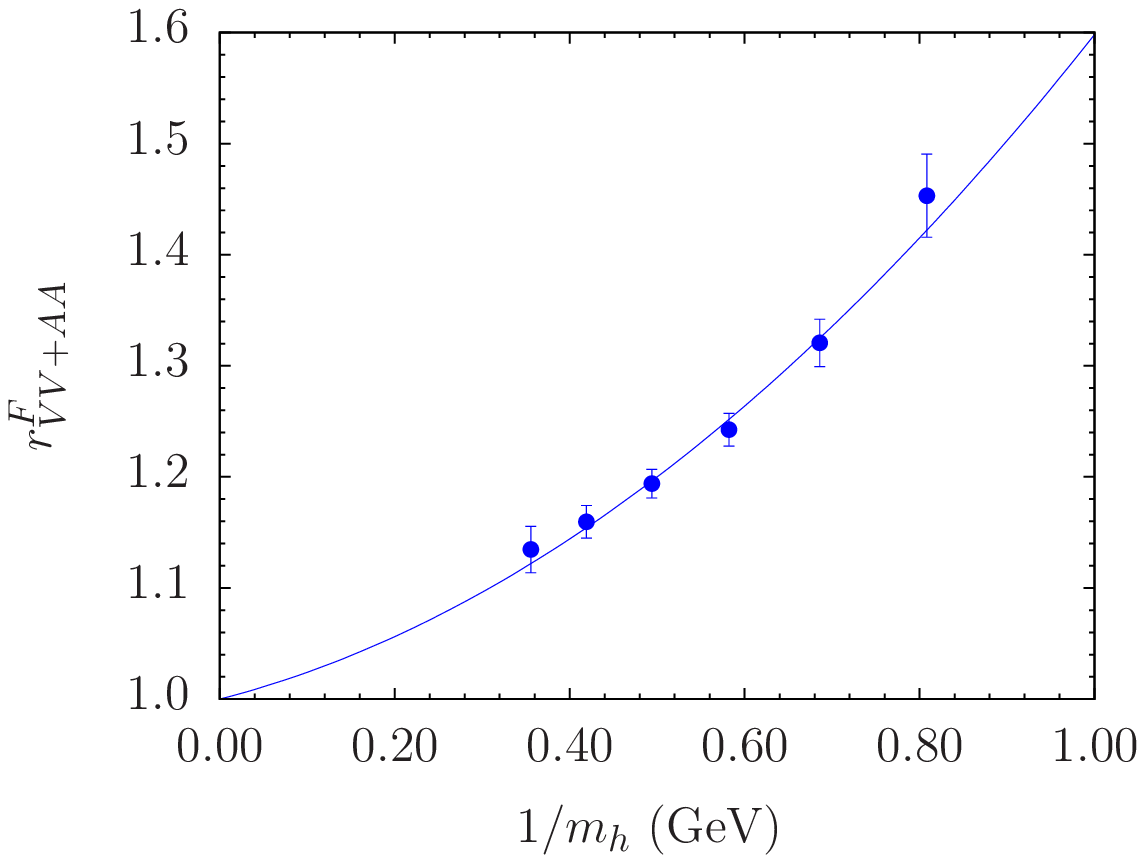}
\end{center}
\vspace{-0.5cm}
\caption{\sl Results for the ratios $r_{VV+AA}$ (left) and
  $r^{F}_{VV+AA}$ (right) as a function of $1/m_h$, where $m_h$ is the
  heavy quark mass renormalized in the $\msb$ scheme at the $\mu=3$
  GeV. The solid lines illustrate the result of a quadratic fit of the
  lattice data and the precisely known value $r^{(F)}_X=1$ in the
  infinite mass limit.}
\label{fig:ratio}
\end{figure}

For the calculation of $B$ and static meson matrix elements we have
applied the ratio method for heavy quarks and optimized smearing
techniques, as discussed in Ref.~\cite{Carrasco:2013zta}. The method
relies on the construction of suitable ratios with exactly known
static limit and an interpolation between the lattice results
evaluated in the accessible charm region and the infinite mass
point. In the case of interest, the quantities $R^{(F)}_X$ defined in
Eq.~(\ref{eq:RX}) tends to a constant in the infinite mass
limit. Therefore, double ratios as  
\beq
\label{eq:doubleratios}
r^{(F)}_X(m_h)=\frac{R^{(F)}_X(m_h)}{R^{(F)}_X(m_h/\lambda)}
\eeq
are equal to 1 in the static limit up to logarithmic corrections which
can be evaluated in perturbation theory. In practice, with this
method,  the results at the $b$-quark mass are obtained after a
relatively small, typically $N_{s}=9$, number of steps  (the precise
value of $N_{s}$ depends on $\lambda$). By iterating the same
procedure for a much larger value of steps, $N_{s}={\cal O}(40)$, one
reaches numerically an asymptotic result which corresponds to the
static limit. Two examples of the lattice data for the double ratios
in Eq.~(\ref{eq:doubleratios}), namely $r_{VV+AA}$ and
$r^{F}_{VV+AA}$,  and of their interpolation to heavier quark masses
are shown in Fig.~\ref{fig:ratio}. 

\begin{figure}[!t]
\begin{center}
\subfigure[]{\includegraphics[width=0.40\textwidth]{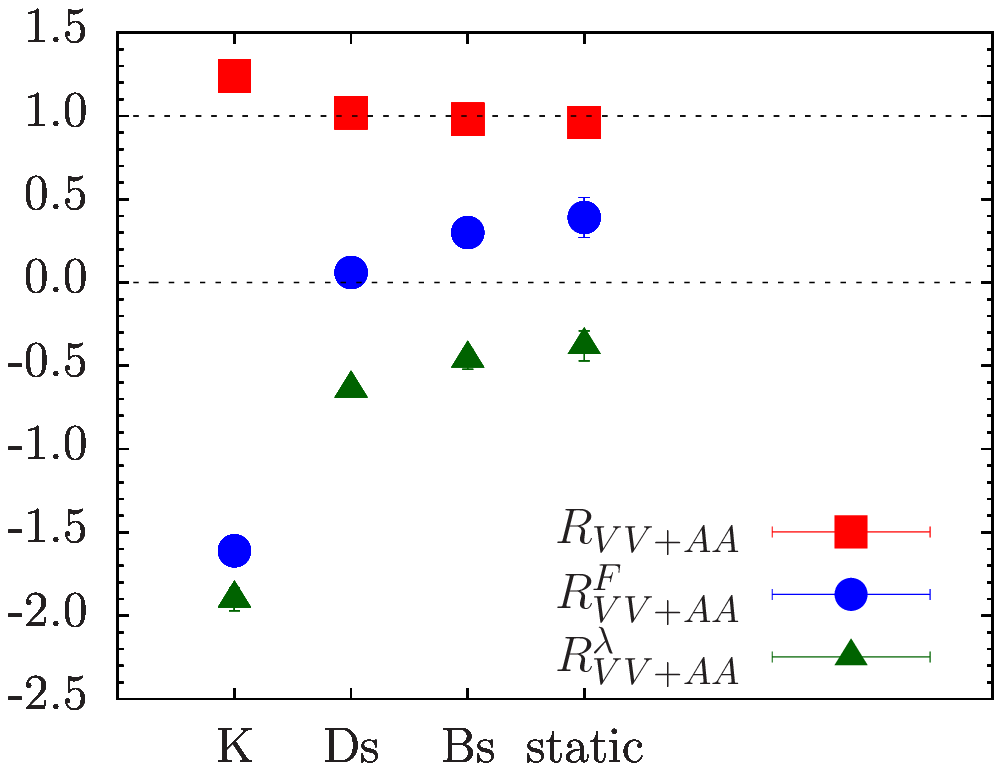}} \hspace{1cm}
\subfigure[]{\includegraphics[width=0.40\textwidth]{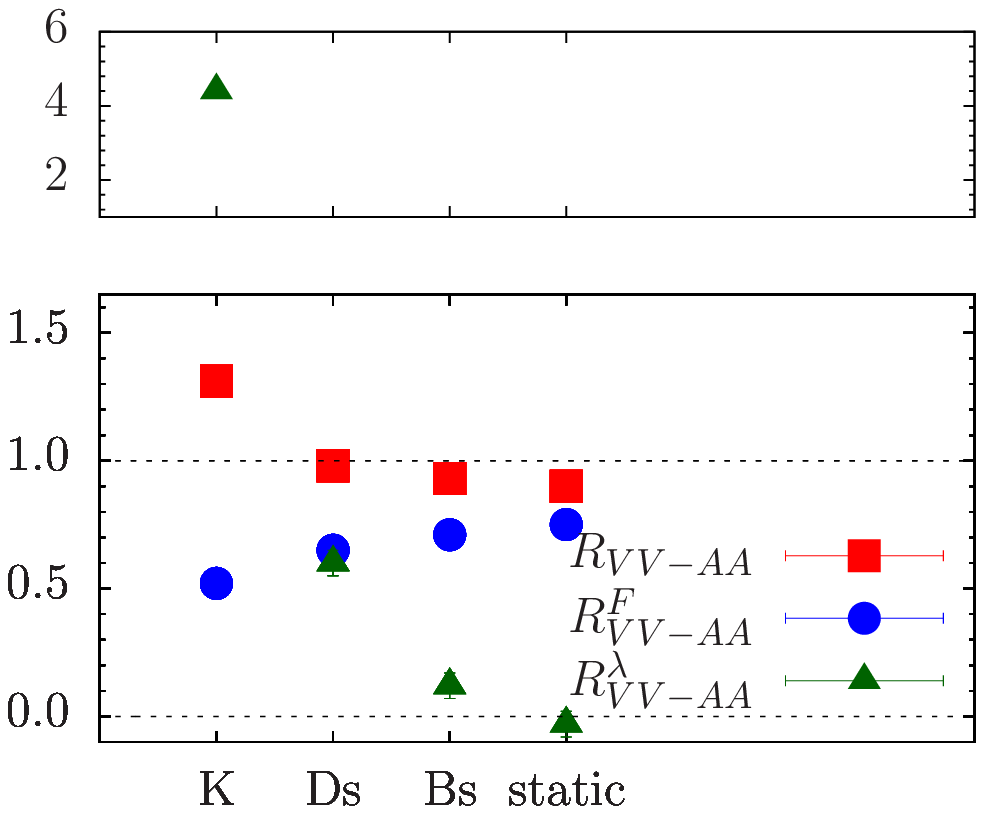}}
\subfigure[]{\includegraphics[width=0.40\textwidth]{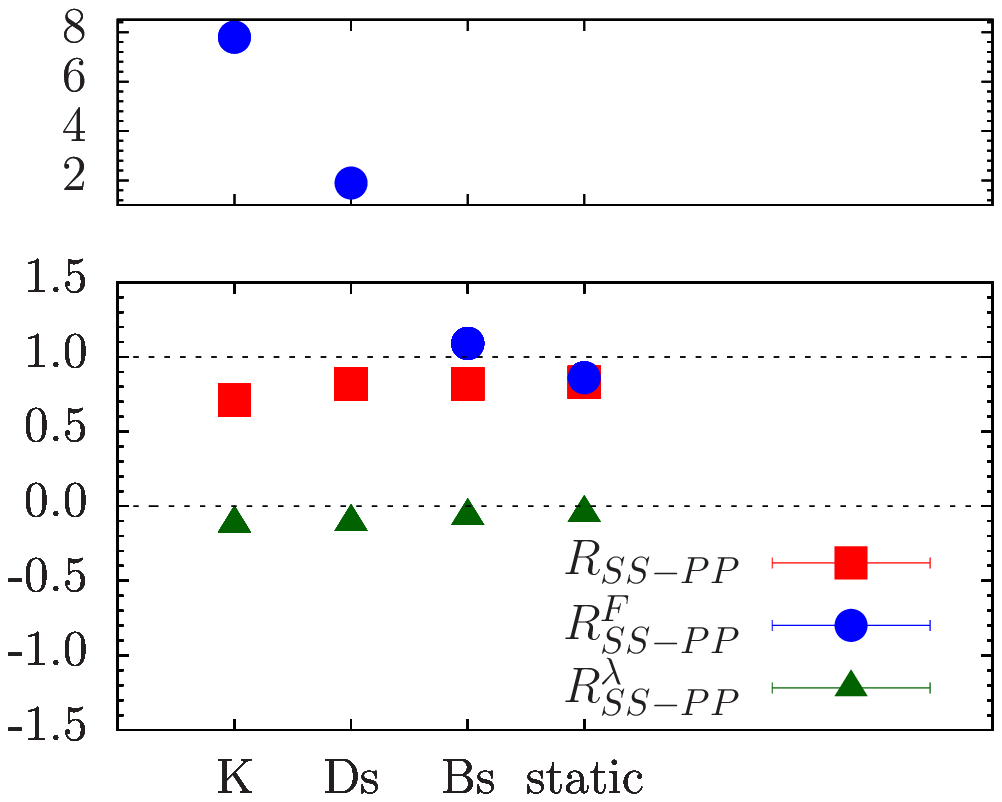}} \hspace{1cm}
\subfigure[]{\includegraphics[width=0.40\textwidth]{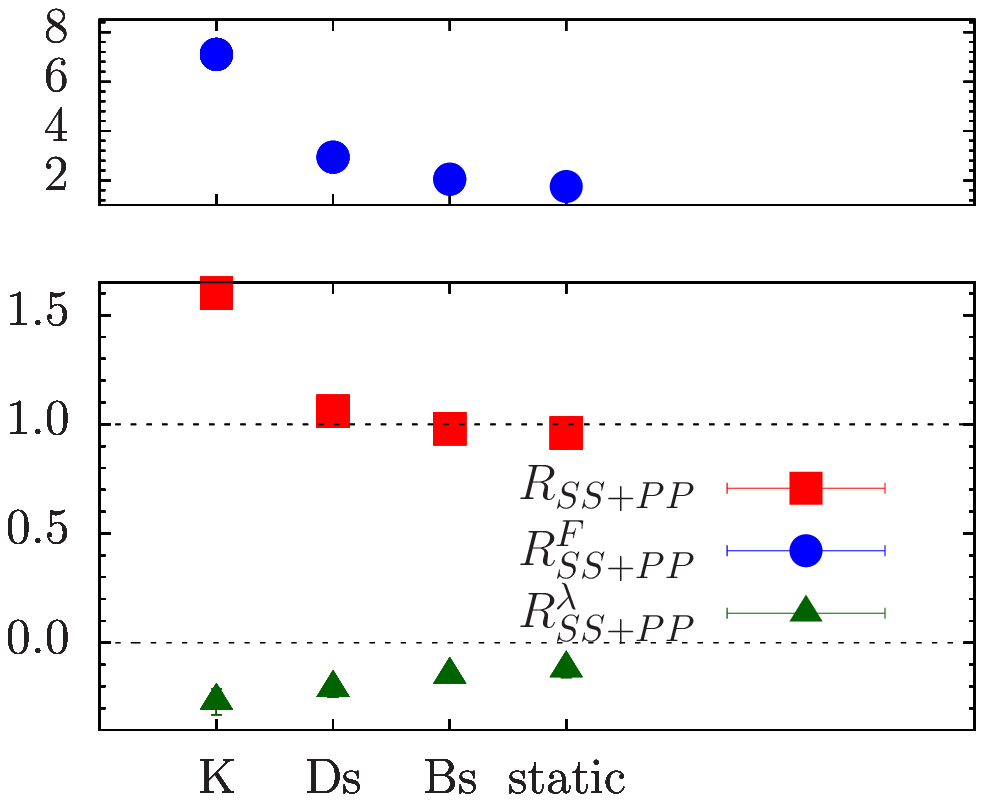}}
\subfigure[]{\includegraphics[width=0.40\textwidth]{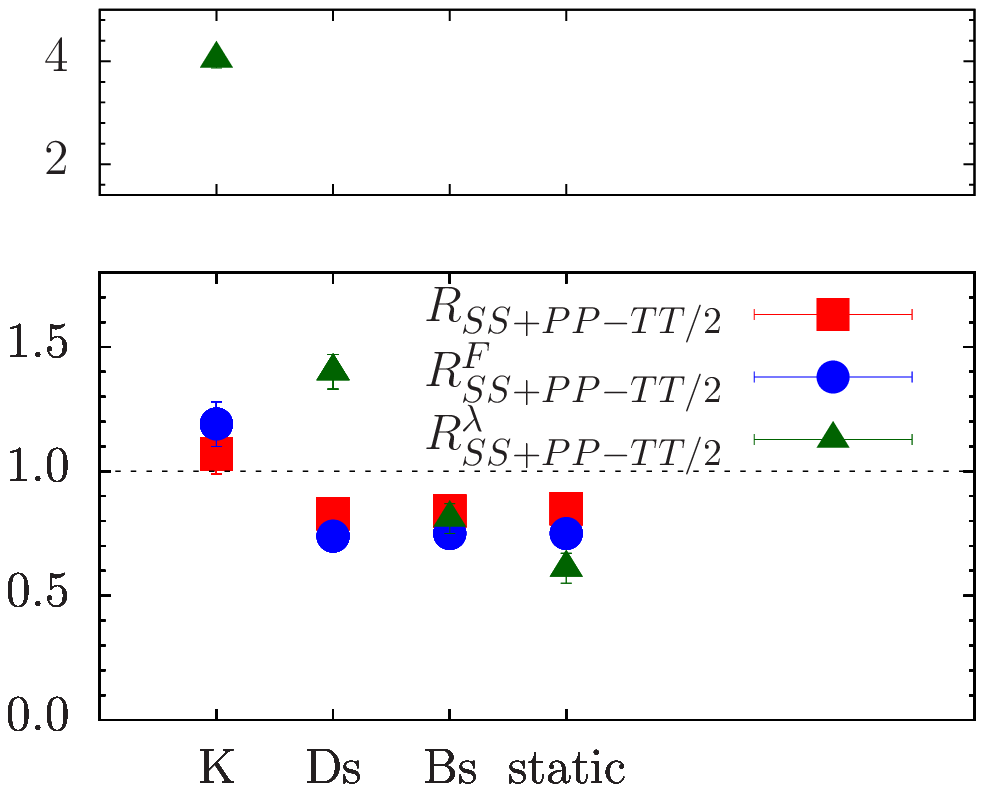}}
\end{center}
\caption{\sl Comparison among the values of the ratios
  $R^{(F,\lambda)}_X$ in the $K$, $D_s$, $B_s$ and static meson
  sectors. The results are renormalized in the $\msb$ scheme of
  Ref.~\cite{Buras:2000if} at the scale $\mu=3$ GeV. For clarity, the
  y-axis in figures (b), (c), (d) and (e) have been splitted. }
\label{fig:comparison}
\end{figure}

Our final results in the continuum limit for the ratios between QCD
and VIA matrix elements in the $K$, $D_s$, $B_s$ and static meson
sectors are collected in Table \ref{tab:results} and shown in
Fig.~\ref{fig:comparison}. The four-fermion operators are renormalized
in the $\msb$ scheme of Ref.~\cite{Buras:2000if} at the scale $\mu=3$
GeV. As a cross check of the calculation, we verified that, by
properly combining the results for the connected and disconnected
matrix elements given in Table \ref{tab:results}, we are able to
reproduce the results for the bag parameters $B_i$ obtained in
Refs.~\cite{Bertone:2012cu,Carrasco:2014uya, Carrasco:2013zta}.
\begin{table}[!h]
\begin{center}
\renewcommand{\arraystretch}{1.3}
\begin{tabular}{lcccc}
\hline  \hline & $K$ & $D_s$ & $B_s$ & static limit  \\ \hline \hline
$R_{VV+AA}$ & 1.24(04) & 1.02(03) & 0.98(03) & 0.96(04)\\
$R_{VV-AA}$ & 1.31(05) & 0.98(03) & 0.93(03) & 0.90(03)\\
$R_{SS-PP}$ & 0.71(03) & 0.82(03) & 0.82(04) & 0.83(04)\\
$R_{SS+PP}$ & 1.60(05) & 1.06(04) & 0.98(04) & 0.96(03)\\
$R_{SS+PP-TT/2}$ & 1.07(08) & 0.83(05) & 0.84(05) & 0.85(05)\\ 
\hline
$R_{VV+AA}^{F}$ & -1.61(08) & 0.06(02) & 0.30(09) & 0.39(12)\\
$R_{VV-AA}^{F}$ & 0.52(04) & 0.65(04) & 0.71(05) & 0.74(05)\\
$R_{SS-PP}^{F}$ & 7.8(4) & 1.89(08) & 1.09(06) & 0.86(06)\\
$R_{SS+PP}^{F}$ & 7.1(2) & 2.94(11) & 2.05(09) & 1.75(10)\\
$R_{SS+PP-TT/2}^{F}$ & 1.19(09) & 0.74(06) & 0.75(05) & 0.78(05)\\
\hline 
$R^\lambda_{VV+AA}$ & -1.90(07) & -0.64(02) & -0.46(06) & -0.38(09)\\
$R^\lambda_{VV-AA}$ & 4.3(2) & 0.60(05) & 0.12(05) & -0.03(05)\\
$R^\lambda_{SS-PP}$ & -0.13(03) & -0.11(03) & -0.07(03) & -0.05(03)\\
$R^\lambda_{SS+PP}$ & -0.27(06) & -0.21(04) & -0.15(03) & -0.12(04)\\
$R^\lambda_{SS+PP-TT/2}$ & 4.04(16) & 1.40(07) & 0.81(06) & 0.61(06)\\
\hline \hline

\end{tabular}
\renewcommand{\arraystretch}{1.0}

\par\end{center}
\caption{\sl Results for the ratios $R^{(F,\lambda)}_X$ between the
  matrix elements of the four fermion operators in QCD and in the VIA,
  defined in Eqs.~(\ref{eq:RX}) and (\ref{eq:Rlambda}). The ratios are
  renormalized in the $\msb$ scheme of Ref.~\cite{Buras:2000if} at the
  scale $\mu=3$ GeV.}
\label{tab:results}
\end{table}

\section{Conclusions}
The results collected in Table \ref{tab:results} and presented in
Fig.~\ref{fig:comparison} show large violations of the VIA,
particularly for the connected contributions in the kaon sector. The
ratios $R^F_{SS+PP}$ and $R^F_{SS-PP}$ in this sector are found to be
as large as $7\div 8$, while the ratio $R^F_{VV+AA}$ has negative
sign, as anticipated in Ref.~\cite{Boyle:2012ys, Lellouch:2011qw}. The deviations from
the VIA decrease significantly, however, as the meson mass
increases. The ratio $R^F_{SS-PP}$, for instance, becomes compatible
with 1 at the $B$ meson mass region, and the ratio $R^F_{VV+AA}$
changes its sign around the charm mass region. Large deviations from
the VIA in the kaon sector are observed also for the matrix elements
of the four-fermion operators with the octet color structure, and some
of the ratios $R^\lambda_X$, which are expected to vanish in the VIA,
are found to be much larger than 1. As in the case of the color
singlet operators, these deviations become significantly smaller going
towards the static limit.

Results similar to those presented in Table \ref{tab:results} and
Fig.~\ref{fig:comparison} are obtained when the operators are
renormalized in the $\msb$ scheme at the scale of 2 GeV or in the
RI-MOM scheme at the same scales. The scheme dependence is an ${\cal
  O}(\alpha_s)$ effect and all our conclusions about the accuracy of
the VIA remain qualitatively valid.  The numerical results in Table \ref{tab:results} and
Fig.~\ref{fig:comparison} are obtained in the $N_f=2$ theory and do not account
for the dynamical sea quark effects of the strange and heavier quarks. However, preliminary ETMC results
with $N_f=2+1+1$ dynamical sea quarks indicates that the systematic effect due to the 
partial quenching do not change the qualitative conclusions described here. We also notice that once the
results for the ratios $R_X$ and $R_X^F$ are combined in order to
reconstruct the values of the B-parameters for the $\Delta F=2$
operators, the large violations of the VIA observed particularly in
the kaon case cancel to a large extent. The B-parameters for the five
independent operators turn out to be of order one (see
Refs.~\cite{Bertone:2012cu,Carrasco:2014uya, Carrasco:2013zta}).

\section*{Acknowledgements}
We warmly thank Petros Dimopoulos and Cecilia Tarantino for useful
discussions and for comments on the manuscript. The research leading
to these results has received funding from the MIUR (Italy) under the
contract PRIN 2010-2011 and from the European Research Council under
the European Union's Seventh Framework Programme (FP/2007-2013) / ERC
Grant Agreements n.~279972 and 267985.

\end{document}